\documentclass[showpacs,twocolumn,twoside,amsmath,amssymb,prl,superscriptaddress,longbibliography]{revtex4-1}
\newcommand*{\balancecolsandclearpage}{%
  \close@column@grid
  \clearpage
  \twocolumngrid
}
\usepackage{times}
\usepackage{graphicx}
\usepackage{amssymb}
\usepackage{url}
\usepackage{epstopdf}
\usepackage[unicode]{hyperref}
\hypersetup{
   unicode=true,          
   plainpages=false,
   colorlinks=true,       
   citecolor=blue,        
}
\begin{document}

\title{Center-of-mass motion as a sensitive convergence test for variational multi-mode quantum dynamics}

\author{Jayson G. Cosme}

\affiliation{New Zealand Institute for Advanced Study, Dodd-Walls Centre for Photonics and Quantum Technology, Centre for Theoretical Chemistry and Physics, Massey University, Private Bag 102904, North Shore, Auckland 0745, New Zealand}

\author{Christoph Weiss}

\affiliation{Joint Quantum Centre (JQC) Durham--Newcastle, Department of Physics, Durham University, Durham DH1 3LE, United Kingdom}

\author{Joachim Brand}

\affiliation{New Zealand Institute for Advanced Study, Dodd-Walls Centre for Photonics and Quantum Technology, Centre for Theoretical Chemistry and Physics, Massey University, Private Bag 102904, North Shore, Auckland 0745, New Zealand}



\date{\today}                                           

\newcommand{\sech}{{\mathrm{sech}}}

\begin{abstract}
%

Multi-mode expansions in computational quantum dynamics promise convergence toward exact results upon increasing the  number of modes. Convergence is difficult to ascertain in practice due to the unfavourable scaling of required resources  for many-particle problems and therefore a simplified criterion based on a threshold value for the least occupied mode function is often used. Here we show how the separable quantum motion of the center of mass can be used to sensitively detect unconverged numerical multi-particle dynamics in harmonic potentials.
Based on an experimentally relevant example of attractively interacting bosons in one dimension, we demonstrate that the simplified convergence criterion fails to assure qualitatively correct results. Furthermore, the numerical evidence  for the creation of two-hump fragmented bright soliton-like states presented by Streltsov \emph{et al.} [PRL 100, 130401 (2008)] is shown to be inconsistent with exact results. Implications for our understanding of dynamical fragmentation in attractive boson systems are briefly discussed.


%
\end{abstract}

\maketitle

Recent advances in the field of ultracold atoms 
have made it possible to observe the quantum dynamics of few to many particles under unitary time evolution \cite{Bloch2008}. 
The opportunity to explain and  predict novel effects
motivates  computational approaches, which face 
the challenge of vast complexity
\cite{He2012,daley2004}. 
Variational multi-mode dynamics seeks to reduce the computational complexity by expanding the wave function with a  small number $M$ of optimised mode functions \cite{Meyer1990,Beck2000,Caillat2005,Faust2010}. 
Specifically adapted for bosonic particles is the multi-configurational time-dependent Hartree method for bosons (MCTDHB) \cite{AlonEtAl2008}. 
It provides a hierarchy of approxmations beyond the Hartree or Gross-Pitaevskii mean-field theory \cite{pitaevskii03:book},
to which it reduces for $M=1$. The ability to represent fragmented Bose-Einstein condensates and correlated wave functions for $M>1$ is the defining feature of the approach. While the limit of large $M$ formally recovers the multi-particle Schr\"odinger equation, it is often impossible to verify convergence through increasing $M$ due to prohibitive computational requirements. This motivates the  search for independent convergence checks.
 
Here we test the convergence of MCTDHB simulations by exploiting the artificial coupling of the center-of-mass (COM) and relative motion in the truncated multi-mode expansion.
In harmonic external potentials and homogeneous gauge fields, the COM dynamics of a many-particle system is independent of the particle interactions by the generalised Kohn's theorem  \cite{Kohn1961,Gibbons2011}. 
Including time-dependent, anisotropic, rotating, or absent trapping potentials of any number of spatial dimensions, this result covers a wide range of experimentally relevant scenarios, where the exact  quantum mechanical time evolution of the COM can be easily obtained. 
Since a convergent simulation is typically required to reproduce the exact COM dynamics, a comparison between both results serves as a sensitive convergence test.

An interesting scenario for quantum dynamics with ultra-cold atoms is provided by attractive bosons in narrowly confining elongated traps, where bright matter-wave solitons of $10^2$ to $10^4$ atoms have been observed \cite{KhaykovichEtAl2002,StreckerEtAl2002,MedleyEtAl2014,McDonaldEtAl2014,NguyenEtAl2014,MarchantEtAl2015}. Fragmentation of the Bose-Einstein condensate can be anticipated from theoretical arguments \cite{CastinHerzog2001}, even though experiments have been largely consistent with Gross-Pitaevskii ($M=1$) theory.
The tendency to form many-particle bound states \cite{McGuire1964}, which are themselves well approximated by the Hartree approximation \cite{CalogeroDegasperis1975}, further motivates the use of multi-mode expansions, and several MCTDHB-based studies have been published \cite{Streltsov2008,StreltsovEtAl2008b,StreltsovEtAl2011}. In this work we find a pathologically slow convergence of the MCTDHB expansion for untrapped or weakly-trapped attractive bosons where the COM length scale becomes of the same order or larger than the typical length scale of relative motion. Specifically, we  find that predictions for the dynamical creation of the two-humped, two-fold fragmented states of attractive bosons named ``fragmentons'' in Ref.~\cite{Streltsov2008}  were based on unconverged MCTDHB simulations and are inconsistent with the exact COM dynamics. 
We further find that previously proposed internal convergence checks of MCTDHB fail to reliably detect unconverged results, including the popular strategy of 
setting a threshold for the smallest eigenvalue of the  single-particle density matrix  to estimate the relevance of the least important mode \cite{Meyer1990,Beck2000,Cao2013}.

For definiteness, we consider the dynamics of $N$ bosons of mass $m$ in one dimension with the Hamiltonian
\begin{equation}\label{ham1}
 \hat{H} = \sum_{i=1}^N h(x_i,t) + g(t)\sum_{i<j} \delta(x_i - x_j) = \hat{H}_R + \hat{H}_r,
\end{equation}
where $h(x,t) = -\frac{\hbar^2}{2m}\frac{\partial^2}{\partial x^2} + \frac{1}{2}m\omega(t)^2 x^2$, and ${g}<0$ is the coupling parameter of attractive interactions \cite{olshanii98}.
Due to the harmonic trapping potential the problem is separable and the COM Hamiltonian $H_R = -\frac{\hbar^2}{2 N m}\frac{\partial^2}{\partial R^2} + \frac{1}{2}N m\omega(t)^2 R^2$ formally defines a single-particle problem in the COM coordinate  $R=N^{-1}\sum_{i}x_i$. The Hamiltonian of relative motion $H_r$ depends only on the $N-1$ distances between particles and commutes with $H_R$. Thus the time evolution of any observables that are purely related to the COM coordinate is completely independent of the interaction strength. This is very useful for checking the convergence of multi-mode simulations.

\emph{Multi-mode quantum dynamics} --  
The MCTDHB method is based on the variational ansatz for the quantum state 
\begin{align}\label{mctdhbeq}
 |\Psi(t) \rangle 
 &=\sum_{n_1, \dots, n_M} C_{n_1,\dots,n_M}(t) \prod_{k=1}^M \frac{1}{\sqrt{n_k!}}[\hat{b}^{\dagger}_k(t)]^{n_k}|\mathrm{vac}\rangle,
\end{align}
with $N=\sum_{k=1}^M n_k$ particles. Both the expansion coefficients and the single-particle functions
$\phi_k(x,t)=\langle x|\hat{b}^{\dagger}_k(t)|\mathrm{vac}\rangle$ are time dependent
and their evolution equations follow from a variational principle (for details see Ref.~\cite{AlonEtAl2008}). The main parameter determining the accuracy and computational effort of
MCTDHB simulations is the number of single-particle modes $M$.
The COM variance  $\sigma^2_R\equiv\langle (R- \langle R \rangle)^2\rangle$
can be obtained from \eqref{mctdhbeq} through the two-body density matrix $\rho^{(2)}(x,y)= \langle \hat{\psi}^{\dagger}(x)\hat{\psi}^{\dagger}(y)\hat{\psi}(x)\hat{\psi}(y) \rangle$ as  \cite{Klaiman2015},
\begin{align}\label{eq:varCOMrho}
 \sigma^2_{R}(t) =  \int \frac{x^2 + (N-1)xy}{N^2(N-1)} \rho^{(2)}(x,y;t)\, dx d y .
\end{align}
Since the expansion \eqref{mctdhbeq} refers to single-particle quantities rather than the separated COM and relative coordinates, it does not trivially respect the separability. The simulated time evolution of the COM variance will thus be exact in two limits: When the expansion  \eqref{mctdhbeq} is fully converged, or when particle interactions vanish ($g=0$). In the latter case the MCTDHB time evolution reduces to uncoupled single-particle Schr\"odinger equations, which can be solved accurately within the chosen discretisation scheme  \cite{AlonEtAl2008}. A simple convergence test is thus obtained by re-running a given simulation with $g=0$.
If the interacting simulation is fully converged, the resulting time evolution of the COM must agree in both cases \footnote{In the Supplementary Information we show explicitly that the time evolution of $\sigma^2_{R}(t)$ is independent of interactions \cite{Supp2016}}. 

\begin{figure}[ht!]
\includegraphics[width=0.95\columnwidth]{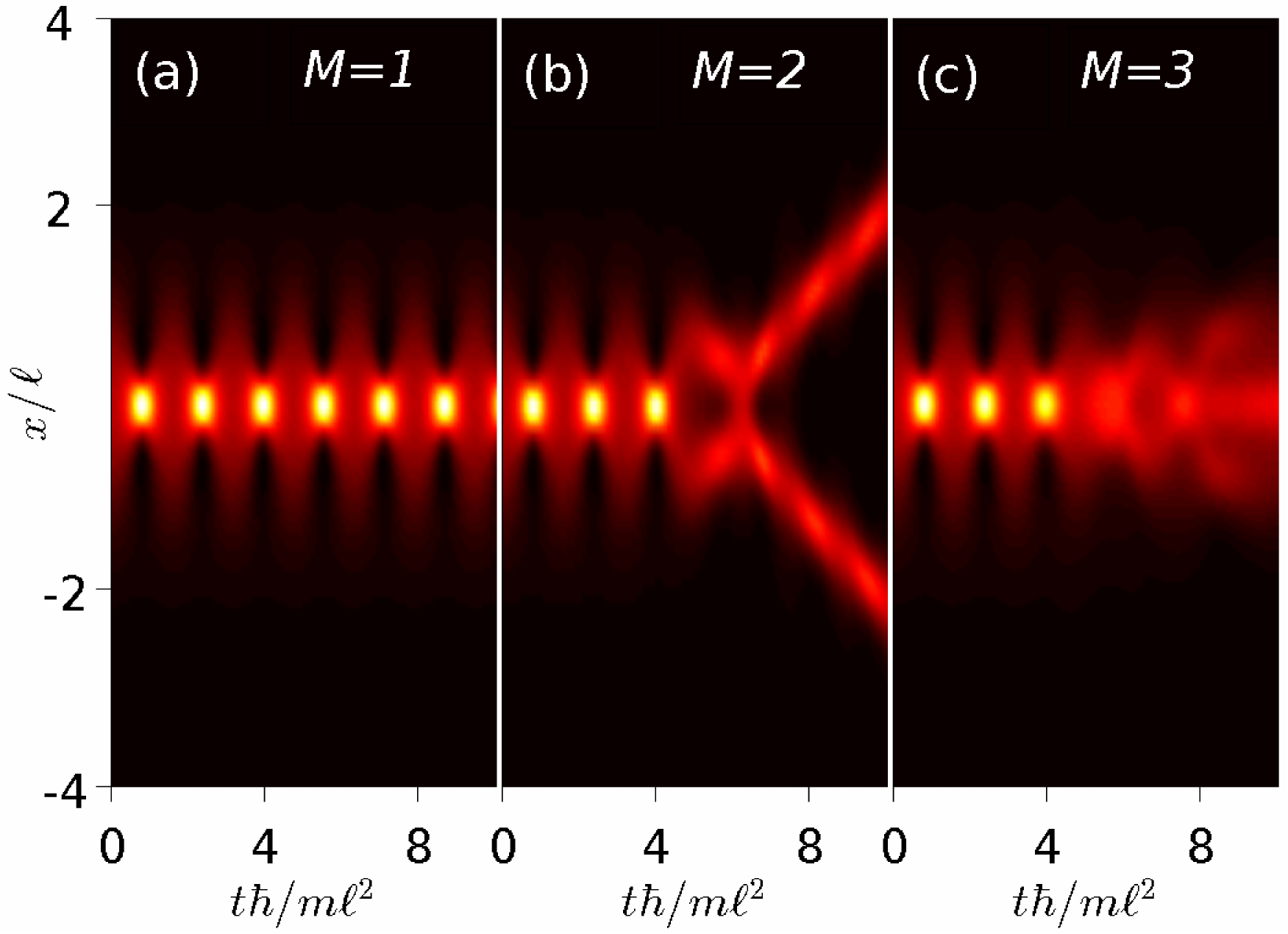}\\
\includegraphics[width=0.95\columnwidth]{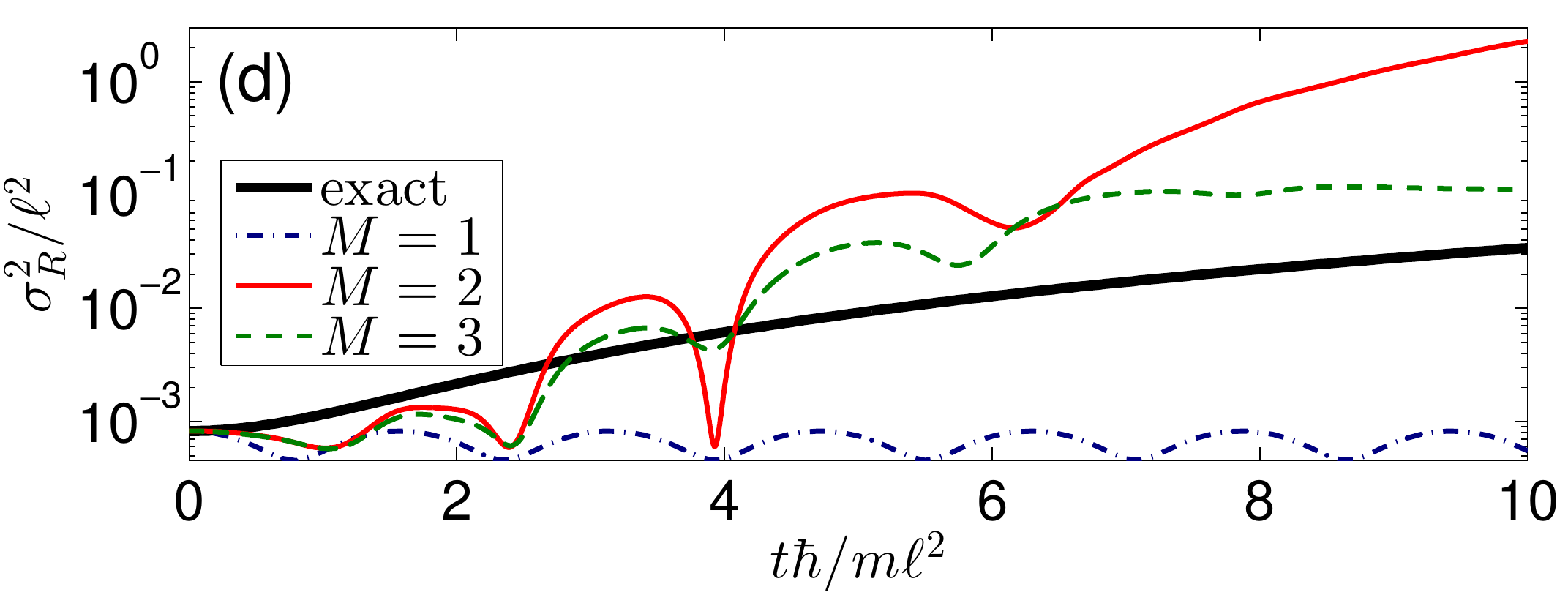}
\caption{(Color online) Time evolution of $N=1000$ attractive bosons prepared in a product initial state corresponding to a mean-field soliton following Ref.~\cite{Streltsov2008}. (a) -- (c) The time evolution of the single-particle density $\langle\hat{\psi}^\dag(x)\hat{\psi}(x)\rangle$ from MCTDHB simulations  for different mode numbers. The $M=2$ simulation (b) was used in Ref.~\cite{Streltsov2008} as evidence for the dynamical formation of two-humped fragented quantum states called ``fragmentons''. (d) The time evolution of the COM variance is compared to the exact result (thick line) from a $g=0$ simulation.
}\label{fig:fragmenton}
\end{figure}

\emph{Dynamical fragmentation} --
As an example we consider the quantum time evolution of a bright soliton state following Ref.~\cite{Streltsov2008}. The initial state is prepared as a simple product state ($M=1$) of $N=1000$ bosons with $\phi_1(x,0)\propto \sech(x/\ell)$, where $\ell$ is a unit length scale, and the time evolution is simulated with MCTDHB in the absence of a trap [i.e.~$\omega=0$ in Eq.~\eqref{ham1}] and with $gm\ell/\hbar^2=-0.008$. The time evolution diagrams of the single-particle density with $M=1$ and $M=2$ shown in panels (a) and (b) of Fig.~\ref{fig:fragmenton} are consistent with the previously published results (see Fig.~1 case III in  Ref.~\cite{Streltsov2008}). The COM variance shown in Fig.~\ref{fig:fragmenton}(d) deviates strongly  from the exact time evolution  and demonstrates that the MCTDHB results   are unconverged. 

Streltsov \emph{et al.} \cite{Streltsov2008} used the simulation result of Fig.~\ref{fig:fragmenton}(b) as evidence for the dynamical formation of two-hump fragmented states. They argued that the simulation corresponds to an interaction quench where the initial state, which is a Gross-Pitaevskii-level approximation to a bright matter wave soliton, is suddently subjected to increased interactions. While a splitting into two equal-sized fragments or solitons is energetically not possible in the Gross-Pitaevskii equation ($M=1$), a fragmented Fock state of the form $|N/2,N/2\rangle$ with overlapping two-hump functions $\phi_{1/2}$ is energetically allowed and can be described within the expansion \eqref{mctdhbeq} with $M=2$ modes \cite{Streltsov2008}. This argument is consistent with the splitting of the single-particle density into two rapidly parting fragments seen in Fig.~\ref{fig:fragmenton}(b), but the exact time evolution of the COM variance is not. The outward motion of the fragments starting shortly after $t=6\,m\ell^2/\hbar$ goes in hand with a rapid increase of the COM variance as seen by the thin (red) line in \ref{fig:fragmenton}(d), growing to almost two orders of magnitude larger than the exact dynamics of $\sigma_R^2$ at $t=10\,m\ell^2/\hbar$. This leads us to the conclusion that dynamical fragmentation cannot happen in just the way that was described in Ref.~\cite{Streltsov2008}, but it leaves open the question whether other dynamical processes might favour the formation of ``fragmentons''.
The $M=3$ simulation [Fig.~\ref{fig:fragmenton}(c), (d)], which was not available at the time of publication of  \cite{Streltsov2008}, shows significant changes compared to the $M=2$ case and further demonstrates that two modes are not sufficient to describe the exact quantum dynamics. Careful examination of the early-time dynamics of the COM variance in Fig.~\ref{fig:fragmenton}(d) reveals an interesting observation: MCTDHB consistently (for $M=1,2,3$) predicts an initial decrease of the COM variance while the exact $\sigma_R^2$  increases monotonically. This illustrates the artificial coupling of the COM with the contracting relative coordinates in MCTDHB. However, importantly, the graphs for $M=2$ and $3$ are on top of each other until $t\approx1\,m\ell^2/\hbar$.
Without the knowledge of the exact COM dynamics, judging from the observed succession of MCTDHB results under the assumption that the expansion \eqref{mctdhbeq} is convergent, one would come to the erroneous conclusion that the observed contraction of the COM variance at early times was a reliable and converged result. 
The obvious discrepancy with the exact result implies that a conventional convergence check based on observing the absence of change while increasing the mode number $M$ fails in this example. Many more modes would be required to converge the expansion \eqref{mctdhbeq}, which is unfeasible. For this reason it is particularly important to be able to solve the COM dynamics exactly in order to detect these artefacts of the simulation.

\begin{figure}[hb!]
\includegraphics[width=0.49\columnwidth]{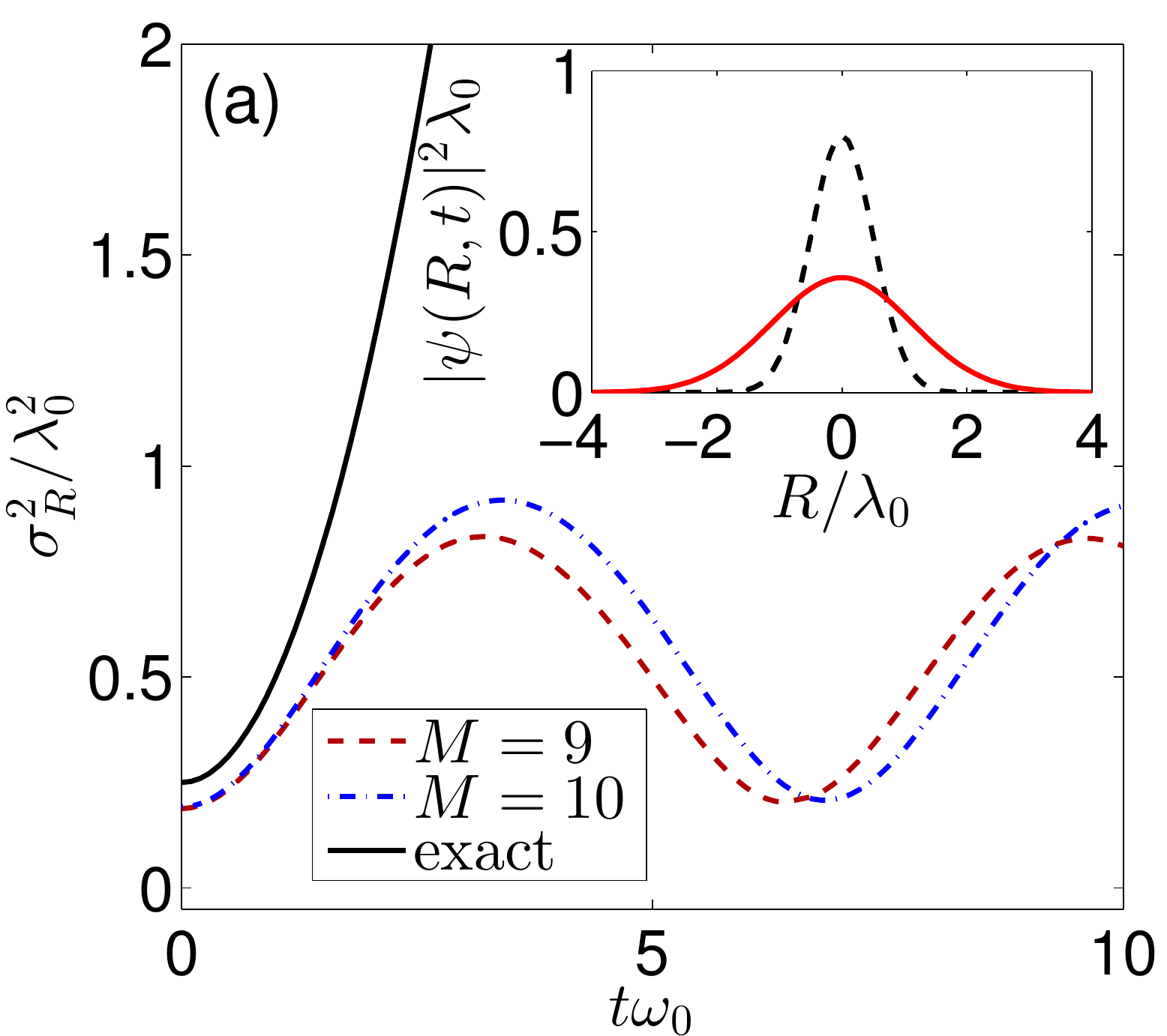} \hfill 
\includegraphics[width=0.49\columnwidth]{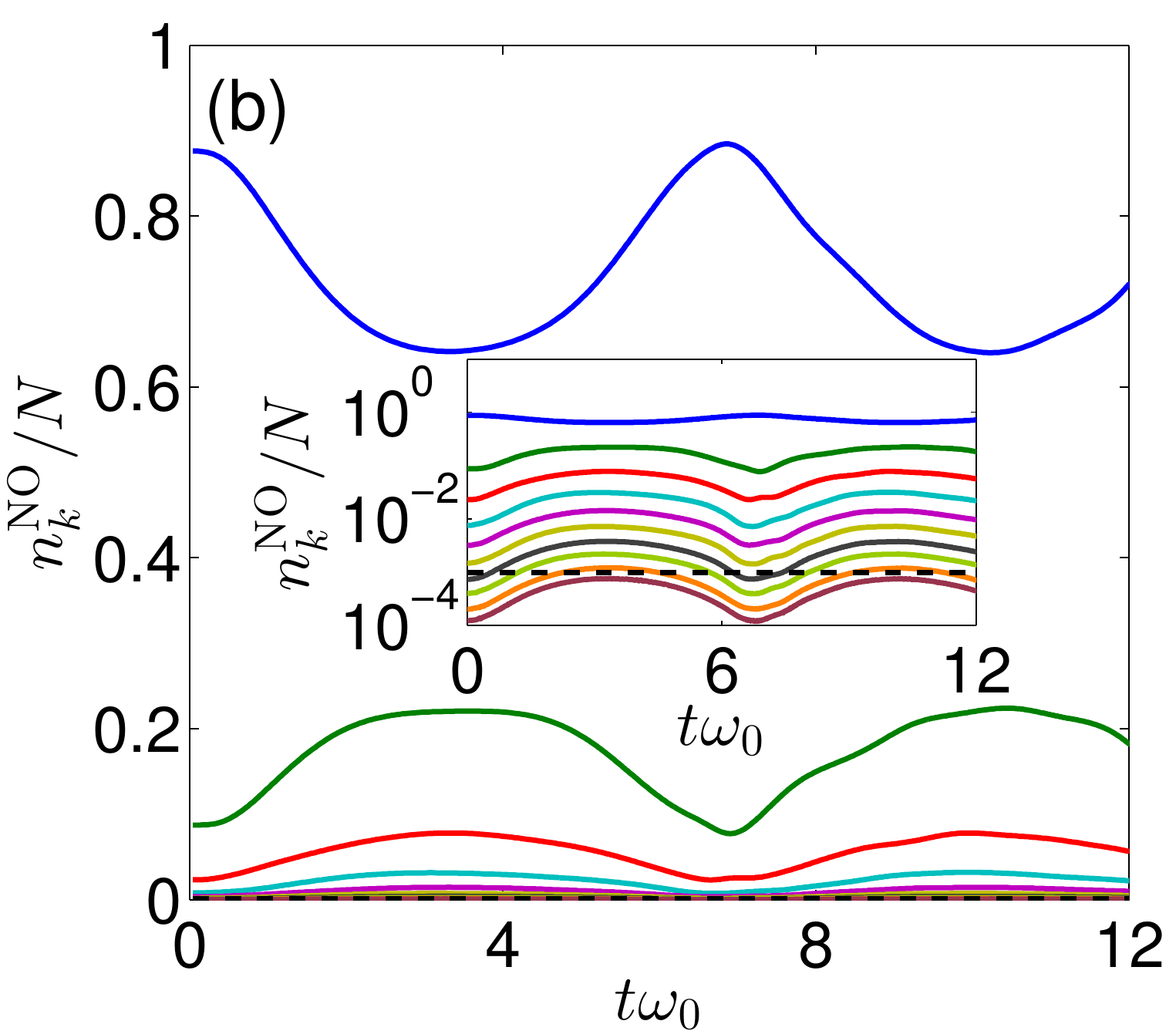}
\caption{Time evolution of $N=2$ particles after sudden release from a harmonic trap; $gm\lambda_0/\hbar^2=-3.16$. (a) Center-of-mass variance exact time evolution
$\sigma_R^2={\lambda_0^2}(2N)^{-1} [1+ ({\hbar t}/{m \lambda^2_0} )^2 ]$ (full line) and  MCTDHB simulation with $M=10$ modes (broken line) showing unphysical breathing oscillations.
Left inset: COM wave function  before ($t=0$, broken line) and after trap release ($t\omega_0=2$, full line).
(b) Eigenvalues of the single-particle density matrix for the MCTDHB simulation.
Right inset: Semi-logarithmic scale showing that the lowest occupancy is below the threshold value $10^{-3}$ for all times.}
\label{fig:comdifg}
\end{figure}

In order to better understand the convergence properties of MCTDHB for attractive bosons we consider a closely related, exactly solvable, and experimentally realisable scenario where two bosons are initially prepared in the ground state of a harmonic trap with frequency $\omega_0$ and released from the trap at $t=0$. 
Simulating the dynamics with up to $M=10$ modes provides a wealth of internal information that can be used to asses the convergence properties of MCTDHB.
Figure \ref{fig:comdifg} shows the results of an unconverged MCTDHB simulation with $M=10$ modes. 

\emph{Natural occupancy criterion} -- The relevance of mode functions in the MCTDHB expansion is  assessed by examining the eigenvalues $n^{\mathrm{NO}}_{k}$ of the single-particle density matrix, also known as natural occupancies, from $\langle\hat{\psi}^\dag(x)\hat{\psi}(y)\rangle = \sum_{k=1}^M n^{\mathrm{NO}}_{k} \varphi^{*}_{k}(x)  \varphi_{k}(y)$, where $\sum_k n^{\mathrm{NO}}_{k}=N$ and eigenvalues are ordered by size $n^{\mathrm{NO}}_{1}\ge  \cdots \ge n^{\mathrm{NO}}_{M} \ge 0$.
A rapidly decreasing sequence of eigenvalues in the exact single-particle density matrix is expected to signal convergence of the multi-mode expansion \eqref{mctdhbeq} \cite{Meyer1990}. Since exact results are usually not available, instead it has become popular to draw conclusions from the natural populations obtained from the variational MCTDHB simulation. A commonly used criterion assumes that the simulation is converged if the smallest relative population lies below a threshold value  \footnote{\label{fn}In fact, near zero eigenvalues are not desirable and may create numerical instabilities, since the MCTDHB algorithm relies on inverting the single-particle density matrix \cite{Beck2000,AlonEtAl2008}} and recently $n^{\mathrm{NO}}_{M}/N < 10^{-3}$ has been used for ultra-cold atom experiments \cite{Sakmann2012,Cao2013,Beinke2015,Mistakidis2015} ($10^{-2}$ in Ref.~\cite{Schurer2015}). The results shown in Fig.~\ref{fig:comdifg} provide an example where the threshold criterion fails, while comparison of the COM dynamics with exact results clearly shows that the simulation is not converged. Beyond the possibility that simply a smaller threshold value may need to be set, we argue  that the logic behind the threshold criterion is flawed because it ignores the possibility that (a) a large number of natural orbitals with very small occupancies can still have an important sum contribution to the density matrix, (b) the natural occupancy of the $M$-th mode
may be underestimated by the variational approach, and
(c) the nonlinear evolution equations of MCTDHB may amplify small inaccuracies in the fractional occupancies into large deviations of observables at later times.  
While good-natured examples were reported in the literature \cite{Manthe1992,Lode2012} where these problems do not arise, all three possibilities play a role in the breakdown of the criterion for attractive bosons. 
Specifically, the variational MCTDHB calculation of the trapped ground state (initial state in  Fig.~\ref{fig:comdifg}) yields a smallest natural occupancy of $n_{10}^\mathrm{NO}/N=1.2\times 10^{-4}$ ($M=10$) compared to the almost four times larger exact value of  $n_{10}^\mathrm{NO}/N=4.5\times 10^{-4}$ (exact), supporting concern (b). It validates point (a) that the cumulative contribution of natural orbitals beyond the 10 highest occupied, $1- N^{-1} \sum_{i=1}^{10} n_i^\mathrm{NO} = 1.4\times 10^{-3}$ (exact), is an order of magnitude larger than the MCTDHB value for the $10^\textrm{th}$ natural occupancy, confirming that the latter is a poor estimate for the former. 
Even though these numbers are several orders of magnitudes smaller than unity and a reasonably faithful representation of the true quantum state might be expected, a  23\% deviation of the COM variance from the exact value indicates a poorly converged result instead.
Finally re-running the MCTDHB simulation with a slightly modified initial state (optimised to $M=9$ modes) we indeed find a sensitive dependence on initial conditions as anticipated in point (c) where a change in the breathing frequency and amplitude will lead to completely different values of $\sigma_R^2$ after a few periods.

\begin{figure}[ht!]
\includegraphics[width=0.85\columnwidth]{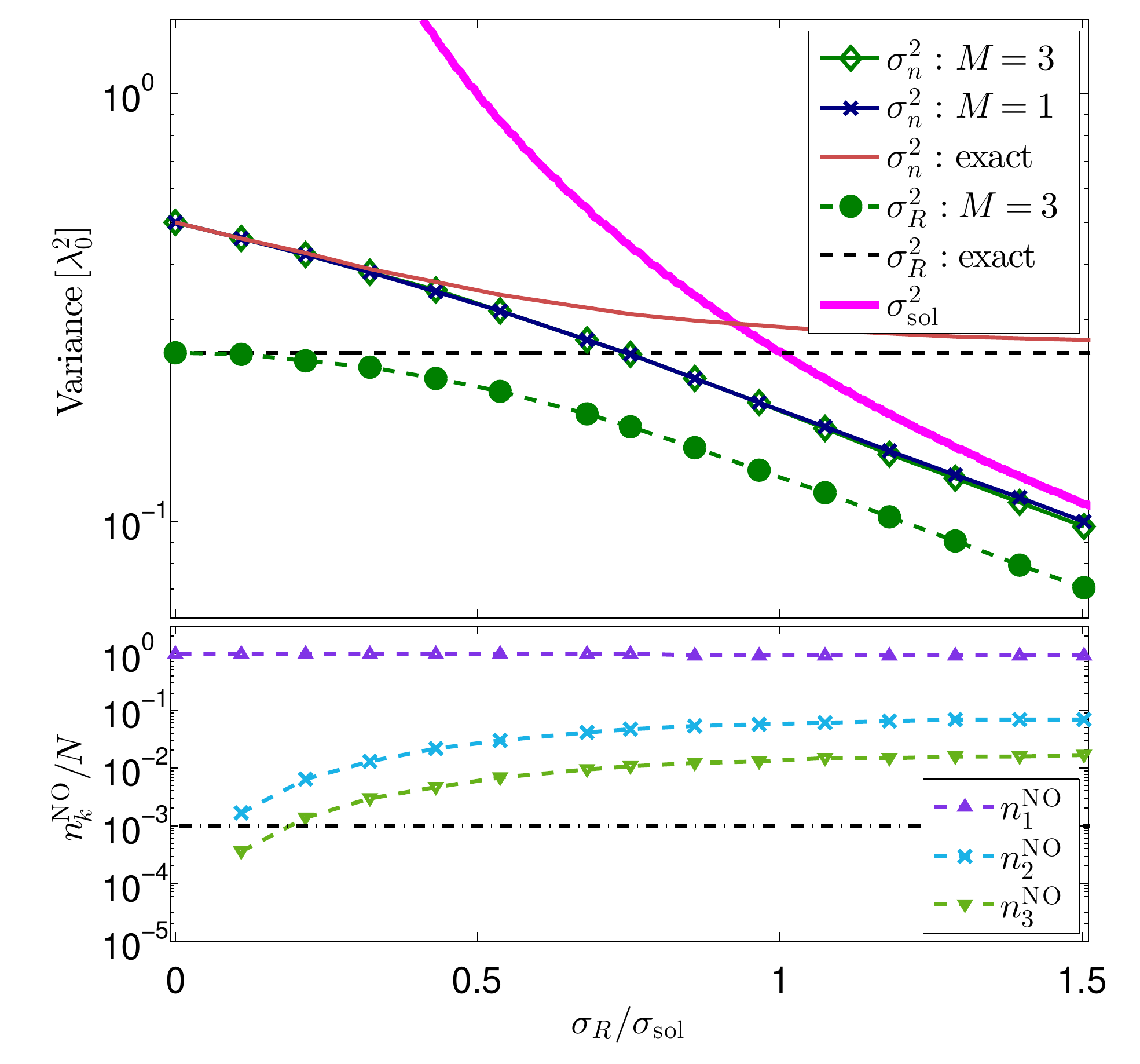}
\caption{Ground state properties of harmonically trapped attractive bosons in one dimension for $N=2$  particles as a function of the length-scale ratio $\sigma_R/\sigma_\mathrm{sol}$.
The top panel shows the COM variance $\sigma_R^2$ and the variance of the single-particle density $\sigma_n^2$ from exact and simulated MCTDHB results.  For comparison also the soliton variance $\sigma_\mathrm{sol}^2$ is shown. This relative motion length scale is clearly seen to influence the variational $M=3$ result in the limit of weak trapping potential.
}
\label{fig:R2r2N2}
\end{figure}

\begin{figure}[t!]
\includegraphics[width=0.85\columnwidth]{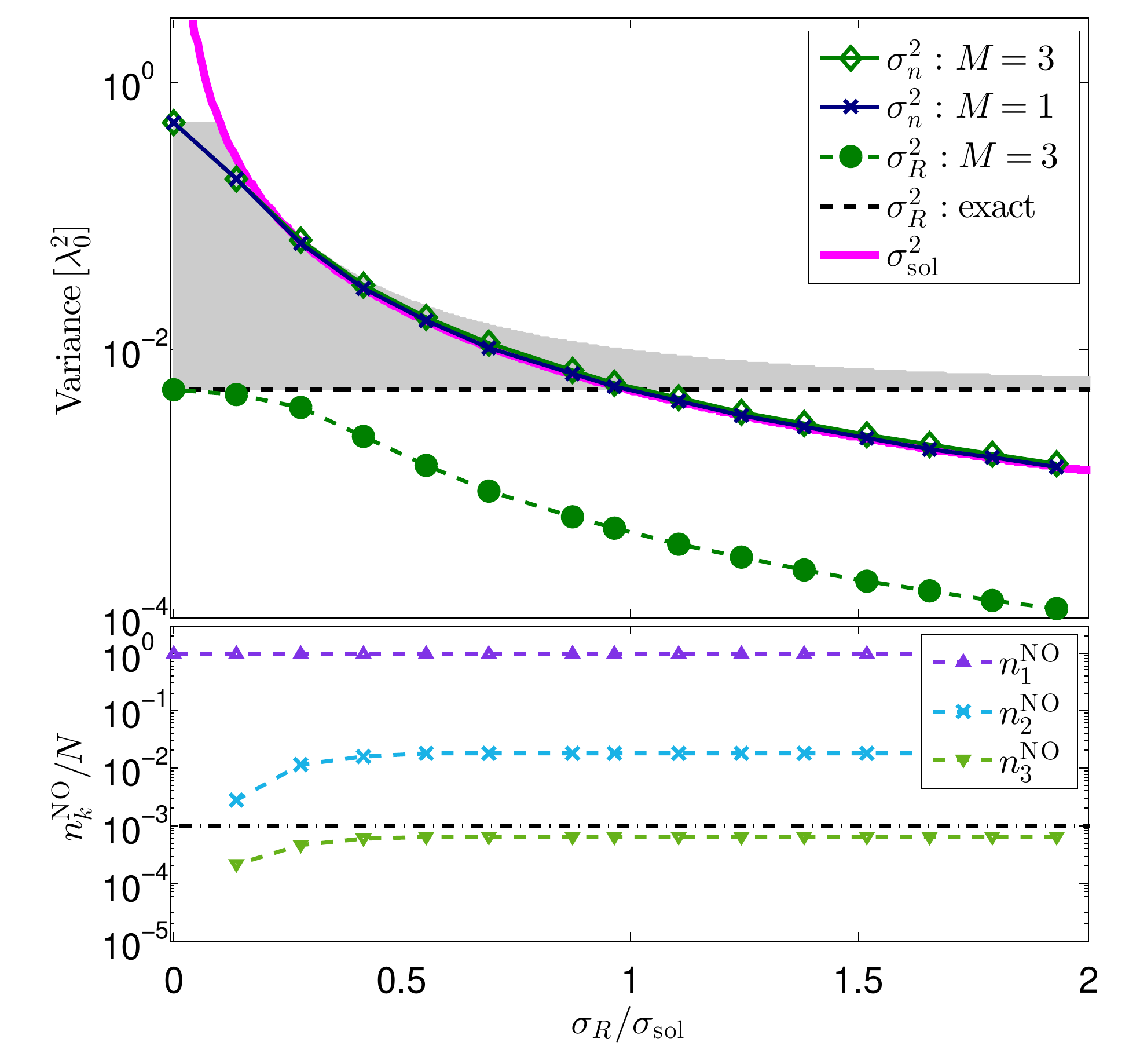}
\caption{Same as Fig.~\ref{fig:R2r2N2} with $N=100$ bosons. The shaded region  depicts the known limits for  $\sigma_n^2$: $\sigma_R^2 \le \sigma_n^2 \lessapprox \sigma_R^2 +\sigma_\mathrm{sol}^2$ \cite{Supp2016}.
}
\label{fig:R2r2N2right}
\end{figure}

\emph{Role of the particle number} -- It is instructive to consider the convergence properties of MCTDHB  in dependence of the available parameters. In contrast to the repulsive Bose gas, which has a dimensionless interaction parameter \cite{lieb63:1}, the interaction strength scales out for untrapped attractive bosons and the only remaining dimensionless parameter is the particle number $N$ \cite{McGuire1964}. 
A second dimensionless parameter  is available in a harmonic trap by comparing relevant length scales.
The  ratio $\sigma_R/\sigma_\mathrm{sol}$ allows for a meaningful comparison of results between varying particle number and is thus used for comparing ground state calculations of attractive bosons in Figs.~\ref{fig:R2r2N2} and \ref{fig:R2r2N2right}.
Here, $\sigma_R^2={\lambda_0^2}/(2N)$ is the ground state COM variance in the harmonic trap and $\sigma_\mathrm{sol}^2= \hbar^4\pi^2/[3 g^2 m^2 (N-1)^2]$ is the variance of the soliton particle density $\propto \sech^2(\pi x/[2\sqrt{3}\sigma_\mathrm{sol}])$ obtained for  the untrapped  ground state with $M=1$ \cite{CastinHerzog2001}, a characteristic length scale determined by particle interactions.

The MCTDHB results for the variance of the single-particle density $\sigma_n^2$ and COM variance $\sigma_R^2$ of Figs.~\ref{fig:R2r2N2} and \ref{fig:R2r2N2right} show good agreement with exact results only for $\sigma_R/\sigma_\mathrm{sol} \ll 1$, which is a weakly-interacting or strong-trap limit where the harmonic potential dominates all length scales of the quantum state.  As soon as the interacting length scale $\sigma_\mathrm{sol}$ becomes comparable to or smaller than the COM length scale, significant deviations from exact results occur for the numerically obtained $\sigma_n^2$ and  $\sigma_R^2$. In the weak-trap regime  $\sigma_R \gg \sigma_\mathrm{sol}$, the deviation can become arbitrarily large.

So could the failure of the MCTDHB approximation be detected by internal criteria, i.e.\ without comparing to exact results? This appears possible for $N=2$ particles, where the threshold of $10^{-3}$ for the lowest occupancy would signal un-converged results for $\sigma_R/\sigma_\mathrm{sol} \gtrsim 0.2$. Inspecting the  sequence of numerical results with increasing $M$ further indicates that convergence is very slow \cite{Supp2016}. The situation is far worse with $N=100$ particles, where increasing $M$ further may not be an option due to limited computational resources \footnote{Indeed, the number of terms in the expansion \eqref{mctdhbeq} is given by the binomial coefficient $\binom{N+M-1}{N}$, which changes scaling from $\sim N$ for $M=2$ to $\sim c^N$ for $M\approx N$, where $c$ depends weakly on $N$ with $2\le c\le 2e$.}. Analysis of the natural occpancies provides the consistent picture of an almost pure Bose-Einstein condensate, with the least occupancy well below the threshold. Further, the main observable $\sigma_n^2$ displays little variation between $M=1$ and $M=3$ on the scale of Fig.~\ref{fig:R2r2N2right} and clearly shows the same trend as function of  $\sigma_R/\sigma_\mathrm{sol}$. We are thus led to the conclusion that the detection of spurious results from MCTDHB is much more difficult and may even be impossible without exact results to compare with, for particle numbers of the order of 100 or larger.

Why is MCTDHB  unable to capture the physics of the weak trap regime while the Hartree approximation ($M=1$) is known to reproduce the exact, untrapped ground state energy to leading order for large $N$ \cite{CalogeroDegasperis1975}, and previous work has found MCTDHB to converge nicely at large $N$ \cite{Lode2012}? The Hartree approximation fails to describe the delocalisation of the COM in the untrapped limit \cite{CastinHerzog2001} because the variational principle, conditioned to minimise the total energy, finds the best compromise in localising the single available mode function $\phi_1(x)$. When a finite number $M>1$ is used in the multi-mode expansion, it is still energetically advantageous to localise the mode functions. Indeed, an infinite number of mode functions is needed to represent a state with delocalised COM but bound relative motion \cite{Supp2016}.

In conclusion, we have demonstrated how easily obtainable, accurate results for the COM variance were useful in detecting unconverged results and in demonstrating the failure of several popular internal convergence checks of MCTDHB. The possible dynamical creation of ``fragmentons'' is re-opened for discussion as numerical evidence in Ref.~\cite{Streltsov2008} turned out spurious. Our findings call for a systematic re-evaluation of the available convergence criteria for numerical quantum dynamics and what is required to claim ``numerically exact'' results \cite{Lode2009,Sakmann2009,Lode2012a,Cao2013,Schurer2015}.
The comparison of COM dynamics with independent, exact results may be useful for other numerical methods of quantum dynamics and is available in any spatial dimensions and for any particle statistics as long as external potentials are at most harmonic.

\begin{acknowledgments}
The data presented in this paper
will be available online~\cite{CosmeEtAl2015Data}.
We thank L.D.\ Carr, B.\ Gertjerenken, M.F.\ Andersen and W.P.\ Reinhardt for discussions. This work has been supported by the Marsden fund of New Zealand (contract number UOO1320) and the  UK Engineering and Physical Sciences Research Council (Grant No. EP/L010844/1).
\end{acknowledgments}


\newpage

\onecolumngrid

\begin{center}
\textbf{\large Supplemental Materials: Center-of-mass motion as a sensitive convergence test for variational multi-mode quantum dynamics}
\vspace{0.6cm}

Jayson G. Cosme$^1$, Christoph Weiss$^2$, and Joachim Brand$^1$\\

$^1${\it New Zealand Institute for Advanced Study, Dodd-Walls Centre for Photonics and Quantum Technology, Centre for Theoretical Chemistry and Physics, Massey University, Private Bag 102904, North Shore, Auckland 0745, New Zealand}

$^2${\it Joint Quantum Centre (JQC) Durham--Newcastle, Department of Physics, Durham University, Durham DH1 3LE, United Kingdom}

\end{center}
\setcounter{equation}{0}
\setcounter{figure}{0}
\setcounter{table}{0}
\renewcommand{\theequation}{S\arabic{equation}}
\renewcommand{\thefigure}{S\arabic{figure}}

\twocolumngrid

\section{Details of the numerical simulations}

Numerical simulations were performed with the open-source QiwiB implementation of MCTDHB \cite{qiwib11}, using standard Runge Kutta time evolution and representing spatial derivatives with five point stencil finite differences. 
For the $N=2$ ground-state results in Figs.~(3) and (4), we have performed simulations on a 1400 point equidistant grid with different values of the coupling constant ${g}$ using $L=30\lambda_{0}$ as the computational box length for $gm\lambda_0/\hbar^2 \ge -0.7794$ and $L=14\lambda_{0}$ for stronger interactions. While for $N=100$, we have used a 2000 point equidistant grid with $L=30\lambda_{0}$ for $gm\lambda_0/\hbar^2 \ge -0.0714$ and $L=10\lambda_{0}$ for stronger interactions. For the time-evolution of the COM variance, we have used 600 point equidistant grid with $L=25\lambda_{0}$ for both $M=9$ and $M=10$ simulations.
We have assured ourselves that the results are converged with respect to changes in these parameters and those of time and space discretization. In addition, we have compared the QiwiB results against an independent implementation of MCTDHB \cite{mctdhblab}, which produced identical results at the reported accuracy. The simulations of the quench dynamics are performed in two steps: (1) relaxation to the ground state of the harmonic trap and (2) time propagation after turning off the trap. 

\section{Interaction dependence of the center-of-mass (COM) variance}

It follows from very general principles that observables linked only to the COM wave function are independent of the interaction strength for at most harmonic potentials or constant gauge fields. Here we demonstrate this explicitly for the COM variance and a many particle system governed by the Hamiltonian of Eq.~(1).
For this purpose we write an arbitrary initial quantum state as 
\begin{equation}
 |\Psi_0\rangle = \sum_{\mu, \nu} c_{\mu\nu} |\chi_\mu\rangle |\Phi_\nu \rangle,
\end{equation}
where $\{|\chi_\mu\rangle\}$ is a complete basis set which depends only on the COM and $\{ |\Phi_\nu \rangle\}$ is a complete basis set which depends on the $N-1$ relative motion degrees-of-freedom. 
We further write the time evolution of the many-body wavefunction as ($\hbar=1$)
\begin{equation}\label{time}
 |\Psi(t)\rangle = e^{-i(\hat{H}_R+\hat{H}_r)t}|\Psi_0\rangle.
\end{equation}
In Eq.~\eqref{time}, the total Hamiltonian was written as $\hat{H} = \hat{H}_R + \hat{H}_r$, where $\hat{H}_R$ is the COM Hamiltonian and the remaining terms, including the interaction-dependent operators, form $\hat{H}_r$. Then we can explicitly write
\begin{align}
  |\Psi(t)\rangle &= e^{-i(\hat{H}_R+\hat{H}_r)t}\sum_{\mu, \nu} c_{\mu\nu} |\chi_\mu\rangle |\Phi_\nu \rangle \\ \nonumber
  &=\sum_{\mu, \nu} c_{\mu\nu}\biggl(e^{-i\hat{H}_Rt}|\chi_\mu\rangle\biggr) e^{-i\hat{H}_rt}|\Phi_\nu \rangle.
\end{align}
The dynamics of the second moment of the COM wave function is obtained as
\begin{align}\label{timecom}
 \langle &\Psi(t)| \hat{R}^2 |\Psi(t)\rangle \\ \nonumber
 &=\sum_{\mu', \nu'}\sum_{\mu, \nu} c^{*}_{\mu'\nu'}c_{\mu\nu} \langle \chi_{\mu'}| e^{i\hat{H}_Rt} \hat{R}^2 e^{-i\hat{H}_Rt}|\chi_\mu\rangle \\ \nonumber
 &\times \biggl( \langle \Phi_{\nu'} |  e^{i(\hat{H}_r-\hat{H}_r)t}|\Phi_\nu \rangle \biggr) \\ \nonumber
 &=\sum_{\mu', \mu, \nu}c^{*}_{\mu'\nu}c_{\mu\nu}\biggl( \langle \chi_{\mu'}| e^{i\hat{H}_Rt} \hat{R}^2 e^{-i\hat{H}_Rt}|\chi_\mu\rangle\biggr) \\ \nonumber.
\end{align}
From the last line of Eq.~\eqref{timecom} it can be seen that the result 
is independent of the interaction strength during the time evolution. The time evolution of the COM variance thus depends only on the initial state (through the expansion coefficients $c_{\mu\nu}$) and the external potential through the COM Hamiltonian $\hat{H}_R$. This fact can be used as a sanity check for MCTDHB simulations, which, if fully converged, should produce the same time evolution for the COM variance for different values of the interaction strength.

\section{Exact solution of the two-particle problem}

The ground state of $N=2$ particles in a time-independent harmonic trap with frequency $\omega_0$ is described by a product of  the center-of-mass (COM) and relative motion wave functions: $\Psi(R,r) = \psi_0(R)\phi_0(r)$. The analytical form of the COM wave function is 
\begin{equation}
 \psi_0(R) = \biggl(\frac{2m\omega_0}{\pi\hbar}\biggr)^{1/4}e^{-m\omega_0 R^2/\hbar}
\end{equation}
where $R=(x_1+x_2)/{2}$. On the other hand, the 
relative motion wave function  with the normalization constant $A$ and harmonic oscillator length scale, $\lambda_{0} = \sqrt{\hbar/m\omega_0}$, can be obtained as
\begin{equation}
 \phi_0(r) = A e^{-r^2/4\lambda^2_{0}}{U}\left(-\frac{\nu}{2},\frac{1}{2},\frac{r^2}{2\lambda^2_{0}}\right) ,
\end{equation}
where $r=(x_2-x_1)$ is the relative coordinate, ${U(a,b,x)}$ is the confluent hypergeometric function of the second kind and $\nu$ comes from the discontinuity in the first derivative due to the delta interaction \cite{BuschEtAl1998}. Explicitly, $\nu$ is calculated by solving the transcendental equation 
\begin{equation}
 \nu = \frac{gm \lambda_{0}}{\hbar^2\sqrt{2}}\frac{\Gamma(1-\nu/2)}{\Gamma(1/2-\nu/2)} .
\end{equation}
The ground-state energy is given by
\begin{align}\label{nuex}
 &E_0^{2,\textrm{exact}}=E_{\mathrm{rel}} + E_{\mathrm{COM}} \\ \nonumber
 &= \hbar\omega_0\biggl(\nu+\frac{1}{2}\biggr) +  \frac{\hbar\omega_0}{2}= \hbar\omega_0(\nu+1),
\end{align}
and
the exact results for the natural occupancy are obtained by numerically diagonalizing of the single particle density matrix
\begin{align}
 \langle\hat{\psi}^\dag(x)\hat{\psi}(y)\rangle &=2\int  dz\Psi^*(x,z,t=0)\Psi(y,z,t=0)  \\ \nonumber
 &=2\int dz \psi_0^*((x+z)/2)\phi_0^*(x-z) \\ \nonumber
&\times \psi_0((y+z)/2)\phi_0(y-z).
\end{align}

After turning off the trap, the Gaussian COM wave function  expands. In particular, the time evolution of the COM wave function represents the textbook example of Gaussian wave propagation \cite{Fluegge1990}
\begin{align}
 \psi&(R,t) \propto \biggl(1+ i\frac{\hbar t}{m {\lambda}^2_{0}} \biggr)^{-1/2} \\ \nonumber
 &\times \mathrm{exp}\biggl(- \frac{R^2}{\lambda^2_{0}[1+i\hbar t/(m\lambda^2_{0})]} \biggr).
\end{align}
The COM wave function spreads leading to a variance increasing quadratically in time
\begin{equation}\label{rmswidth}
 \sigma^2_R(t)=\frac{\lambda_{0}^2}{4}\left[1+ \biggl( \frac{\hbar t}{m \lambda^2_{0}} \biggr)^2\right].
\end{equation}
The relative motion after trap release, on the other hand, is dominated by the bound state of the attractive $\delta$ interactions. Indeed, since the $\delta$ function has exactly one bound state, near the origin the relative motion wave function will approach this  bound state in the long time-limit and possible other contributions from scattering state will disperse. The initial relative motion wave function can be expressed in terms of the bound state and scattering states:
$ \phi_0(r) = c_{\mathrm{b}}\phi_{\mathrm{BS}}(r) + \int dk e^{ikr} c_k$, where the bound state is $\phi_{\mathrm{BS}}(r)  = \sqrt{\frac{{m|g|}}{2\hbar^2}}~\mathrm{exp}\left( \frac{-m|g||r|}{2\hbar^2} \right)$. Then, the expected variance of the relative motion wave function in the long time limit must be $\sigma_r^2 \ge \sigma^2_{\mathrm{BS}} = \int dr r^2 |\phi_{\mathrm{BS}}(r)|^2 = 2 \hbar^4 /m^2{g}^2$.

\section{MCTDHB simulations with $N=2$ particles}
\label{sec:convcheck}

In the main text we have demonstrated the ambiguity of the studying the eigenvalues of the single-particle density matrix, the natural occupations, for a specific MCTDHB simulation. For $N=2$ particles we are able to vary the number of modes $M$ over a good range, which permits a conventional study of convergence with respect to mode number. It is further possible to analyse the shape of the two-particle wave function in the MCTDHB approximation, which sheds some light on the unphysical coupling of relative and COM motion in the truncated multi-mode expansion. 

\subsection{Convergence of MCTDHB results with increasing $M$}

We test the convergence by checking whether relevant quantities, e.g.\ the variational ground-state energy, remain unchanged as the number of modes $M$ is increased.
The results in Table~\ref{tab1} are still varying at the level of several percent between $M=8$ and $M=10$ and thus indicate, correctly, that the MCTDHB expansion  converges very slowly and is not yet fully converged with 10 modes. While this way of testing convergence is reliable and has produced the correct answer, varying $M$ from one to ten modes is a luxury that can only be afforded for small particle numbers $N$. In simulations with hundreds to thousands of particles (e.g.~\cite{Streltsov2008,StreltsovEtAl2011,Schmitz2013}), the options for choosing $M$ are severely limited due to the unfavorable scaling of numerical effort when both $N$ and $M$ are large.

\begin{table}
\renewcommand{\arraystretch}{2}
 \begin{tabular}{ c | c | c | c | c | c | c |}
 \cline{2-7}
  & \multicolumn{3}{|c|}{$\tilde{g}=-3.1623$} & \multicolumn{3}{||c|}{$\tilde{g}=-2$} \\
 \hline
  \multicolumn{1}{|c|}{$M$} & $E/\hbar\omega_0$ & $n_0^\mathrm{NO}/N$ & $n_1^\mathrm{NO}/N$ & \multicolumn{1}{||c|}{$E/\hbar\omega_0$} & $n_0^\mathrm{NO}/N$ & $n_1^\mathrm{NO}/N$\\ \hline
  \multicolumn{1}{|c|}{1} & -0.5787 & 1 & 0 & \multicolumn{1}{||c|}{-0.0915} & 1 & 0 \\ \hline
  \multicolumn{1}{|c|}{3} & -1.1451 & 0.9342 & 0.0536 & \multicolumn{1}{||c|}{-0.1356} & 0.9613 & 0.0316\\ \hline
  \multicolumn{1}{|c|}{5} & -1.3817 & 0.9062 & 0.0701 & \multicolumn{1}{||c|}{-0.2244} & 0.9483 & 0.0392\\ \hline
  \multicolumn{1}{|c|}{8} & -1.5546 & 0.8846 & 0.0825 & \multicolumn{1}{||c|}{-0.2788} & 0.9387 & 0.0451\\ \hline
  \multicolumn{1}{|c|}{10} & -1.6213 & 0.8761 & 0.0872 & \multicolumn{1}{||c|}{-0.3005} & 0.9355 & 0.0470\\ \hline
  \multicolumn{1}{|c|}{Exact} & -1.9527 & 0.8251 & 0.1142 & \multicolumn{1}{||c|}{-0.3993} & 0.9202 & 0.0563\\ 
  \cline{1-4} \cline{5-7}
 \end{tabular}
  \caption{Ground state energy $E$ and the two largest natural occupations from MCTDHB calculations of $N=2$ trapped bosons.
}\label{tab1}
\end{table}

\begin{figure}[htb!]
\includegraphics[width=1\columnwidth]{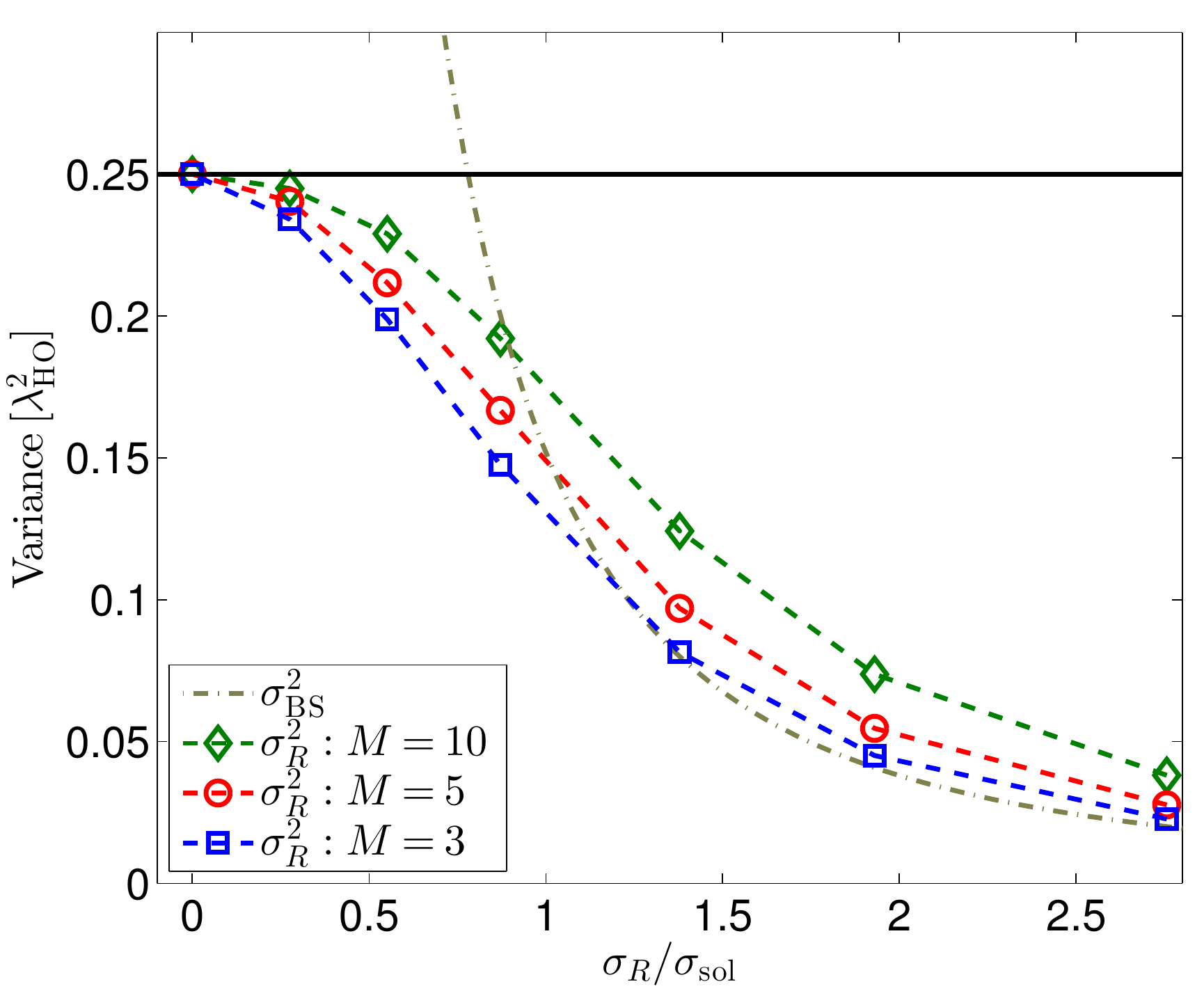}
\caption{COM variance for the trapped ground state as a function of interaction strength from MCTDHB simulations for different values of $M$ (symbols). The solid horizontal  line denotes the exact ground-state COM variance  $\sigma^2_R=\lambda_{0}^2/4$. For comparison, the dash-dotted line shows the exact relative-coordinate variance $\sigma^2_{\mathrm{BS}}$ of the two-particle bound state.}
\label{fig:R2r2}
\end{figure}

We have also considered how the dependence of the COM variance of the trapped ground state on the interaction strength $g$ changes for different numbers of modes $M$. For brevity, we introduce the dimensionless interaction parameter where $\tilde{g}=g m\lambda_{0}/\hbar^2$. For $N=2$, this coupling constant is related to the ratio between relevant length scales via $\tilde{g} = - 0.55~\sigma_R/\sigma_{\mathrm{sol}}$, where the prefactor changes with $N$.
Figure~\ref{fig:R2r2} compares the COM variance from MCTDHB calculations with the exact result $\sigma^2_R=\lambda_{0}^2/4$ from Eq.~(2) of the main text. It is apparent that the MCTDHB results deviate severely from the exact values for strongly attractive interaction, and that convergence of the MCTDHB expansion with increasing the number of modes $M$ is exceedingly slow.

\begin{figure}[ht!]
\includegraphics[width=1\columnwidth]{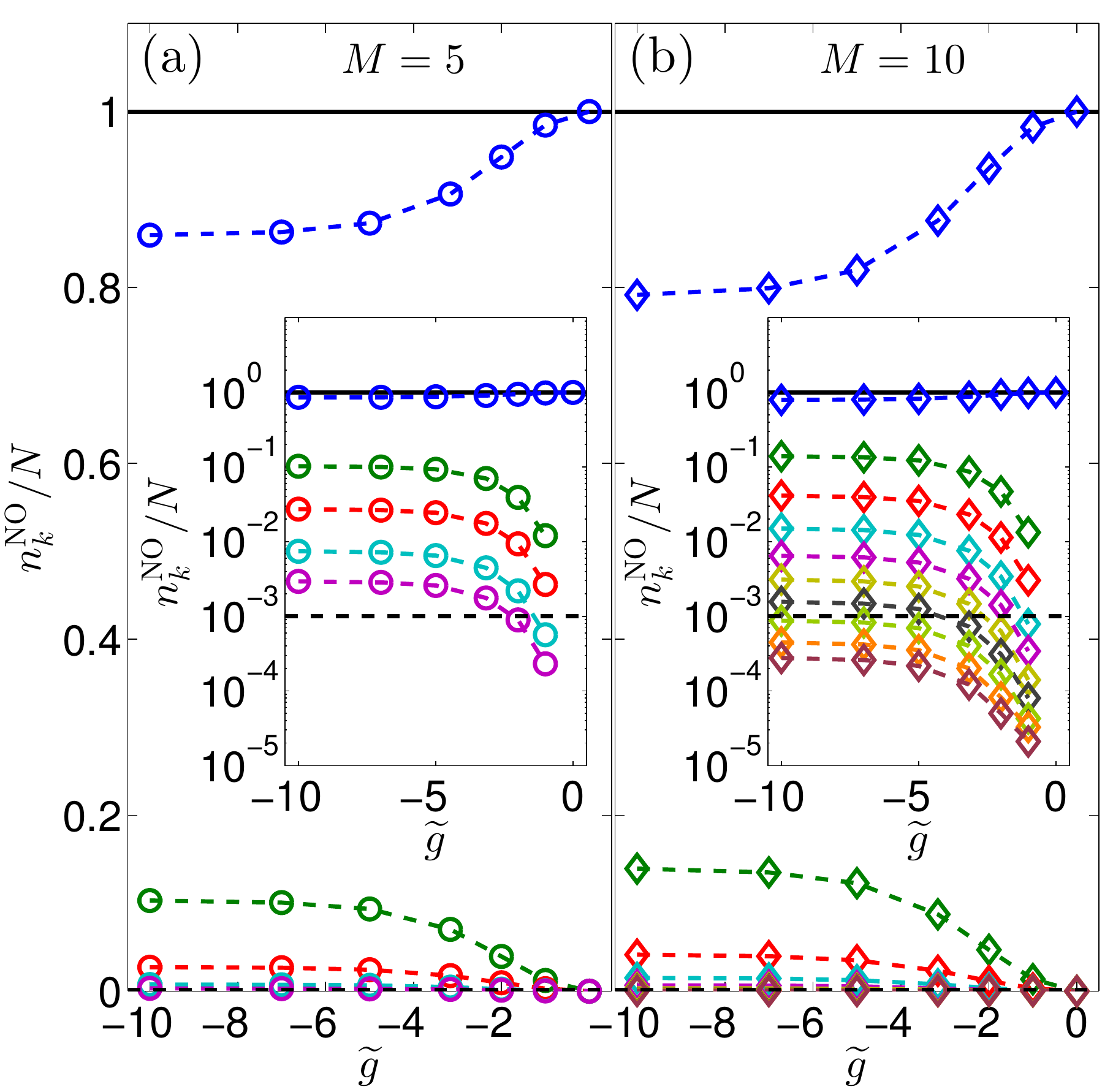}
\caption{(Color online) Natural occupations for the MCTDHB calculations of Fig.~\ref{fig:R2r2}
for (a) $M=5$ and (b) $M=10$. The dashed horizontal line denotes the convergence criterion of $0.1 \% = 10^{-3}$.}
\label{fig:m5m10gspop}
\end{figure}

In Fig.~\ref{fig:m5m10gspop}, we present the dependence of the natural occupation on the interaction parameter. A couple of remarks pointing to the ambiguity of this convergence indicator are in order. First, by looking at the results for $M=10$ and $\tilde{g}=-10$ one is tempted to conclude that MCTDHB has already converged since three orbitals are below $0.1 \%$. But, we know from Fig.~\ref{fig:R2r2} that for the same interaction strength the MCTDHB COM variance is still far from the exact value. Second, it can be seen that all the natural occupations, except for the highest one, are shifted up as the number of single-particle modes $M$ is increased. This further exemplifies the failure of the convergence requirement based on the lowest occupancy. For example, the 5th lowest natural occupation at $\tilde{g}=-2$ is below $0.1\%$ for $M=5$, while this is not true for $M=10$, where the 5th single-particle mode is now above the cut-off value. 

\begin{figure}[htb!]
\includegraphics[width=1.0\columnwidth]{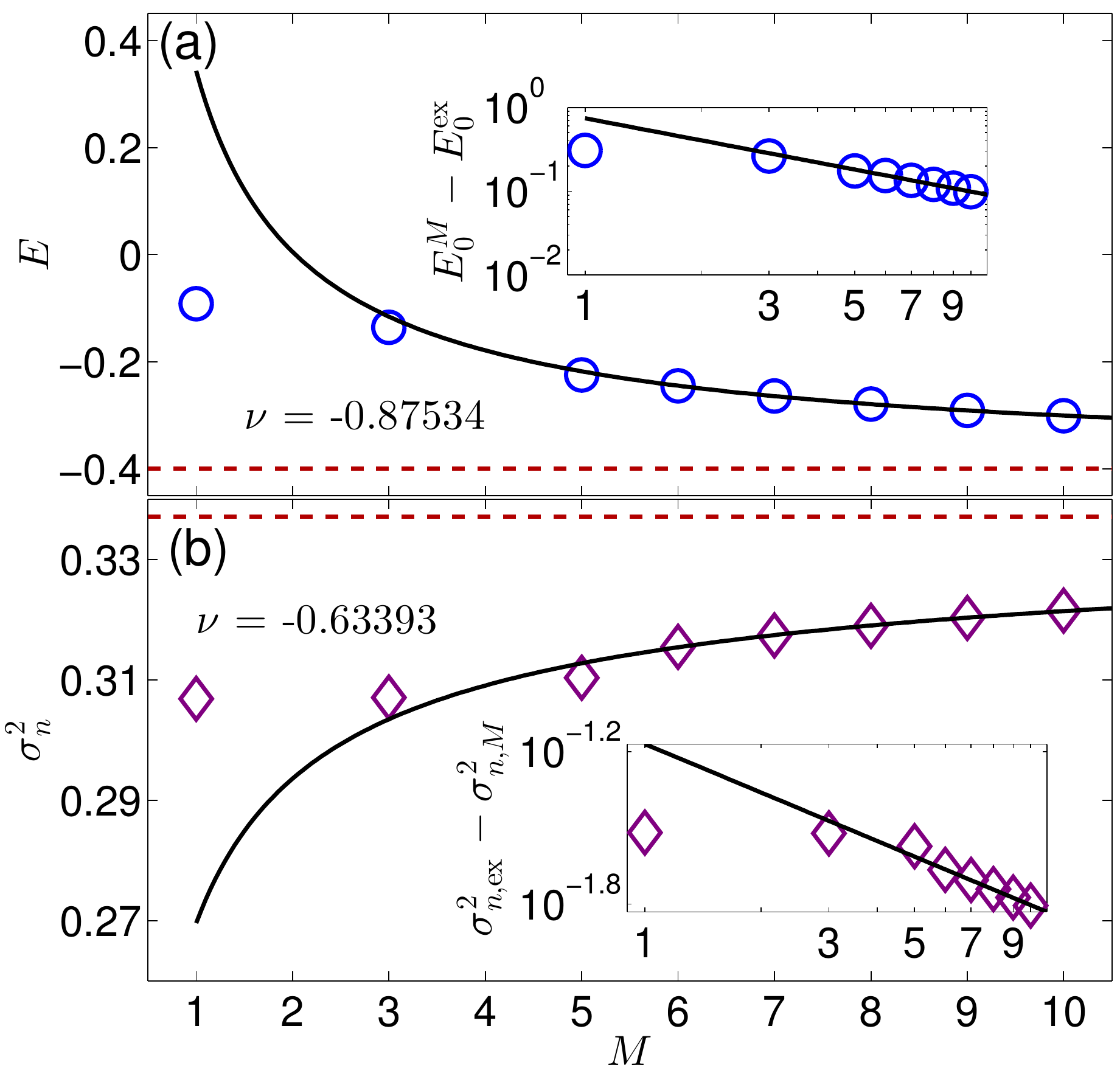}
\caption{Convergence with increasing number of modes $M$ for $\tilde{g}=-2$: (a) Ground-state energy and (b) single-particle density variance. Dashed horizontal lines correspond to the exact values. Symbols denote the MCTDHB results for different $M$. The exponent of the power law fit, $\nu$, is also shown. (Inset) Log-log plot of the absolute difference between the exact and MCTDHB results.}
\label{fig:convM}
\end{figure}

Lastly, we look at the rate of convergence for observables, in particular the ground state energy and the single-particle density variance, as depicted in Fig.~\ref{fig:convM}. We find that the MCTDHB results approach the exact values with a slow power law as a function of the number of single-particle modes $M$, i.e.\ $|O_\mathrm{exact}-O_M| \sim M^\nu$. The empirical exponent $\nu$ lies between $-1$ and $-\frac{1}{2}$, which indicates  slightly faster convergence than the $-\frac{1}{2}$ leading exponent of the full-CI expansion of two 1D bosons with point interaction in a fixed harmonic oscillator basis \cite{Grining2015}. We note  that this power-law behavior only sets in for mode numbers larger than $M=6$ in this case and the convergence rate is significantly slower for smaller $M$. This may be because the variational optimisation of the modes is particularly effective for small $M$. It also means, however, that in the pre-power-law regime of $M<6$ increasing the number of modes brings even less effect than the empirical power law would suggest that governs larger regimes of $M$. Unfortunately, for particle numbers in the hundreds or larger, the scaling of computational resources practically limits the application of MCTDHB exactly to the small $M$ regime.

\subsection{Delocalization of COM: Unphysical coupling of COM and relative motion}

\begin{figure}[ht!]
\includegraphics[width=0.5\columnwidth]{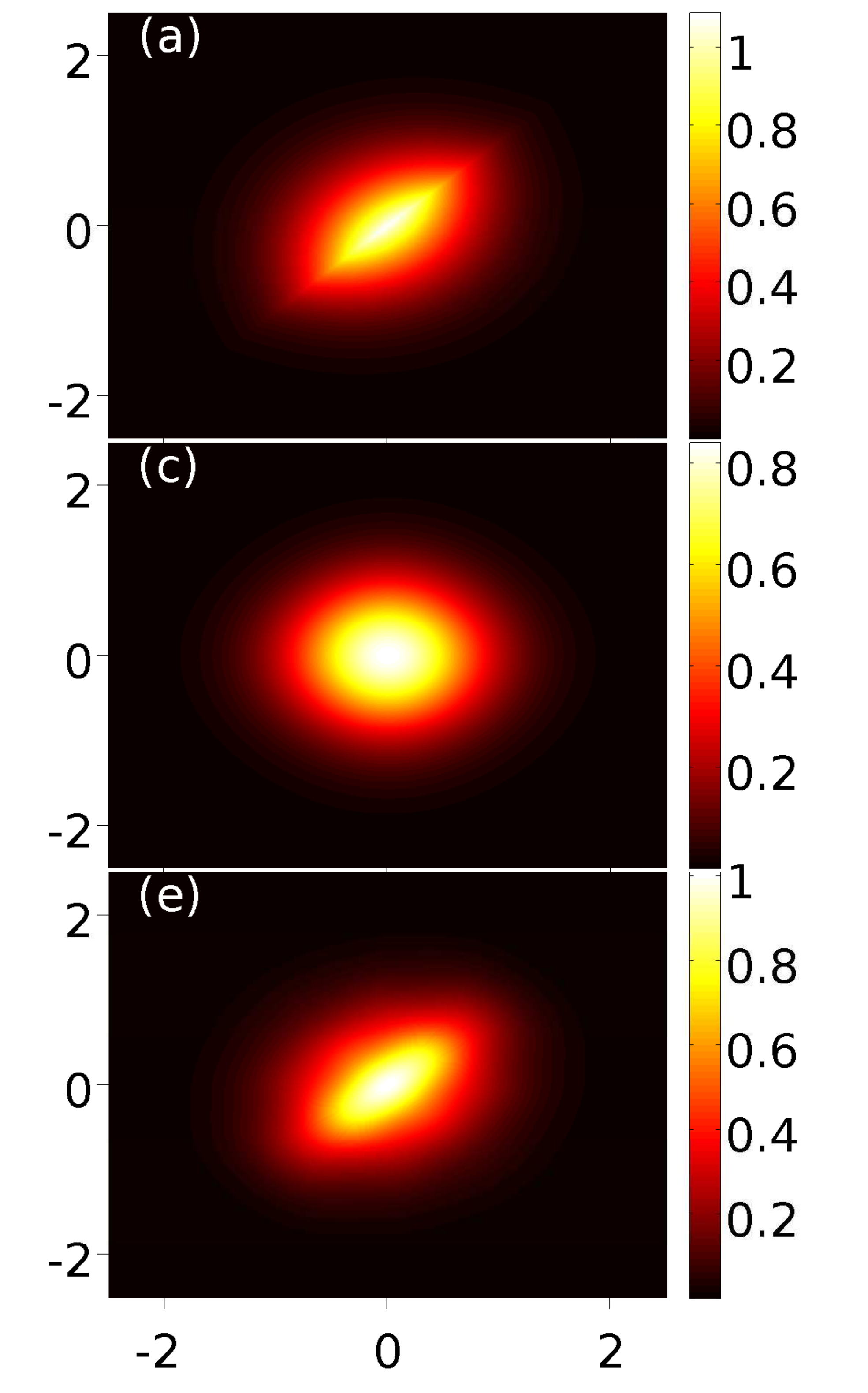}\includegraphics[width=0.5\columnwidth]{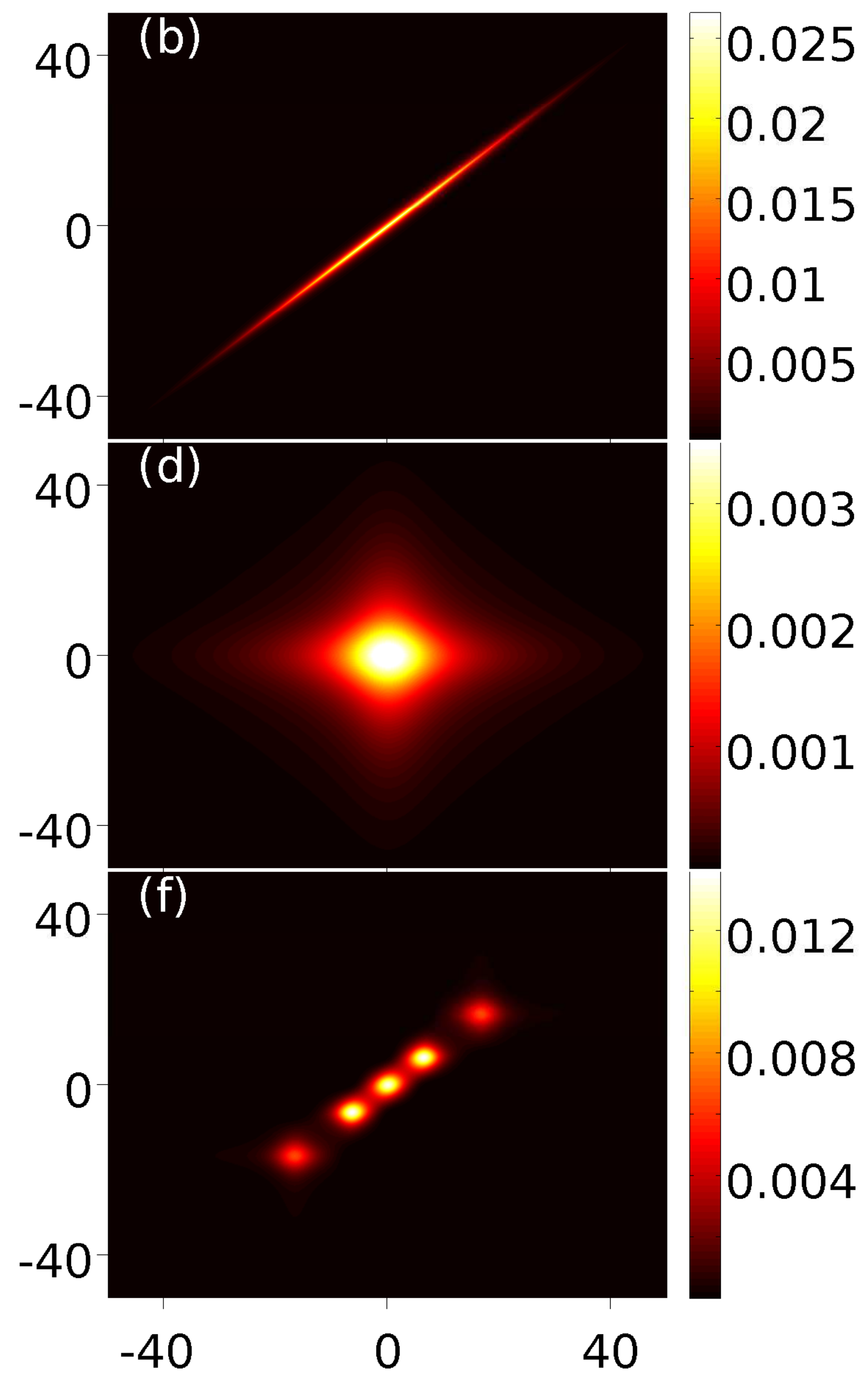}
\caption{Two-body density $\rho^{(2)}(x,y) = 2|\Psi(x,y)|^2$
for $\tilde{g}=-1$ at different times: Left [(a),(c),(e)] $t\omega_0=0$ and right [(b),(d),(f)] $t\omega_0=30$. Comparison between  [(a),(b)] Exact, [(c),(d)] Gross-Pitaevskii ($M=1$), and [(e),(f)] $M=5$.}
\label{fig:2body}
\end{figure}

In order to understand how MCTDHB deals with the competing length scales and why it violates the separation of COM and relative motion, it is instructive to plot the two particle density, as shown in Fig.~\ref{fig:2body} for $N=2$. In these plots, the diagonal ($x=y$) represents the COM coordinate $R$ and the antidiagonal ($x=-y$) the relative motion coordinate $r$. While the left hand panels relate to the ground state of $\tilde{g}=-1$, the right hand panels relate to a later time $t\omega_0=30\,$ after trap release. Here the COM wave function has expanded significantly according to Eq.~\eqref{rmswidth}, whereas the relative-motion bound state is hardly changed. The upper two panels show the exact result and panel (b) clearly demonstrates the diverging length scales. Note the changing spatial scale between panels (a) and (b). 

The middle panels (c) and (d) present the $M=1$ (Gross-Pitaevskii) result. At this level, the COM and relative motion length scales are identical because the product form of the state with a single mode function $|\Psi(x,y)|^2=|\phi(x)|^2|\phi(y)|^2$ together with the inversion symmetry of the problem forces a four-fold symmetry and leaves no option to distinguish the two diagonal directions.  For long times, the wave function expands in both directions and the COM and relative motion length scales are identical.

The lower panels (e) and (f) report an MCTDHB simulation with  $M=5$ modes. The trapped ground state in \ref{fig:2body}(e) is approximated better than with $M=1$, although some more detailed features are missing. The long-time profile in Fig.~\ref{fig:2body}(f) shows five separated peaks with each one exhibiting a four-fold symmetry and resembling the $M=1$ result, albeit on a different scale. 
As the MCTDHB expansion is a sum over symmetrized product states, different numbers of modes $M$ will produce up to $M$ peaks with diagonal -- off diagonal symmetry. Thus for given $M$, the COM and relative motion are strongly coupled and expansion dynamics, where the COM length scale grows over time, will not be captured correctly. Furthermore, this discretised behavior due to finite $M$ leads to an impractical number of $M$ needed to correctly model the limit of a delocalised COM but localised relative motion.

\section{Proof of the bounds for the width of the single-particle density}

In this section, we are going to prove the bounds for the variance of the exact single-particle density manifested in the shaded region in Fig. 4 of the main text. For systems with a Hamiltonian that is separable in the COM and relative motion coordinate, we can conveniently introduce the following change-of-variables for the COM coordinate $R$:
\begin{equation}
 R = \sum_{j=1}^N x_j/N
\end{equation}
and for the relative motion coordinate with $i \geq 2$ \cite{Liu2010}:
\begin{equation}\label{eq:transfo}
 r_i = \sqrt{\frac{i-1}{i}}\biggl(x_i - \frac{1}{i-1}\sum_{k=1}^{i-1}x_k\biggr).
\end{equation}
This means that the many-body wave function is given by
\begin{equation}
 \Psi(x_1,x_2,\dots,x_N) = \chi(R)\Phi(r_2,r_3,\dots,r_N).
\end{equation}
This allows us to write the single-particle density profile as
\begin{equation}\label{eq:spd1}
 \rho(x_N) = \int^{\infty}_{-\infty} \prod_{j=1}^{N-1}dx_j F(R)G(r_2,r_3,\dots,r_N),
\end{equation}
where,
\begin{equation}
 F(R) = |\chi(R)|^2
\end{equation}
and
\begin{equation}
 G(r_2,r_2,r_3,\dots,r_N) = |\Phi(r_2,r_3,\dots,r_N)|^2
\end{equation}
Note that we can express $r_N$ as:
\begin{equation}
 r_N = \frac{N}{\sqrt{N^2-N}}(x_N - R).
\end{equation}
Then we can transform the integration from $\prod_{j=1}^{N-1}dx_j \to |\mathcal{J}|\prod_{j=2}^{N-1}dr_j dR$, where $\mathcal{J}$ is the corresponding Jacobian. The single-particle density becomes
\begin{align}
 \rho(x_N) &= |\mathcal{J}|\int^{\infty}_{-\infty} \prod_{j=2}^{N-1}dr_j dR \;F(R) \\ \nonumber
 &\times G(r_2,r_3,\dots,\frac{N}{\sqrt{N^2-N}}(x_N - R)).
\end{align}
The $(N-2)$ integrations over $dr_j$ can now be done separately and be used to define a new function 
\begin{align}
 H(x_N - R) =& |\mathcal{J}| \int_{-\infty}^{\infty} \prod_{j=2}^{N-1}dr_j \times \\ \nonumber
 &G(r_2,r_3,\dots,\frac{N}{\sqrt{N^2-N}}(x_N - R)),
\end{align}
which allows us to write  the single-particle density as
\begin{equation}\label{eq:spd5}
  \rho(x_N) = \int^{\infty}_{-\infty} dR \;F(R)H(x_N - R).
\end{equation}
One can see that indeed the single-particle density profile is just a convolution between the $|\chi(R)|^2$ and another function ${H}$  that is associated with the relative-motion wave function. 

The function $H$ can be further interpreted as the mean-density for a fixed COM position. This can be justified by following Ref.~\cite{CalogeroDegasperis1975}. We write the single-particle density in this case as
\begin{align}
 \rho&(x'_N|R) = \int^{\infty}_{-\infty} dx_1 \dots dx_N \delta\biggl(R - \sum_{k=1}^N x_k/N\biggr) \\ \nonumber
 &\times \delta(x'_N-x_N) \biggl| \Phi\biggl(\frac{x_2-x_1}{\sqrt{2}},\sqrt{\frac{2}{3}}(x_3 - \frac{1}{2}(x_2+x_1)), \\ \nonumber
 &\dots, \sqrt{\frac{N-1}{N}}x_N - \sqrt{\frac{1}{N^2-N}}\sum_{k=1}^{N-1}x_k\biggr) \biggr|^2.
\end{align}
The integration over $dx_N$ can be easily done due to the presence of the $\delta$-function leading to
\begin{align}
 \rho&(x'_N|R) = \int^{\infty}_{-\infty} dx_1 \dots dx_{N-1} \delta\biggl(R - \sum_{k=1}^{N-1} \frac{x_k+x'_N}{N} \biggr) \\ \nonumber
 &\times \biggl| \Phi\biggl(\frac{x_2-x_1}{\sqrt{2}},\sqrt{\frac{2}{3}}(x_3 - \frac{1}{2}(x_2+x_1)), \\ \nonumber
 &\dots, \sqrt{\frac{N-1}{N}}x'_N - \sqrt{\frac{1}{N^2-N}}\sum_{k=1}^{N-1}x_k\biggr) \biggr|^2.
\end{align}
Again we transform the integration variables using Eq.~\eqref{eq:transfo} such that $\prod_{j=1}^{N-1}dx_j \to |\mathcal{J}|\prod_{j=2}^{N}dr_j$, where we define 
\begin{equation}
r_N = \sqrt{\frac{N-1}{N}}\biggl(x'_N - \frac{1}{N-1}\sum_{k=1}^{N-1}x_k\biggr).
\end{equation}
This yields
\begin{align}
 \rho&(x'_N|R) = |\mathcal{J}|\int^{\infty}_{-\infty} \prod_{j=2}^{N}dr_j \biggl| \Phi\biggl(r_2,\dots, r_N\biggr) \biggr|^2 \\ \nonumber
 &\times \delta\biggl(R - x'_N - r_N\sqrt{\frac{N-1}{N}}\biggr).
\end{align}
Lastly, we integrate over the $dr_N$ to find that indeed $H$ is equal to $\rho(x_N|R)$,
\begin{align}
 \rho(x'_N|R) = H(x'_N-R) 
\end{align}

It is straightforward to show that the variances add in a convolution provided that at least one of the functions is centered at the origin (in our case the COM wave function):
\begin{align}\label{eq:spdineq}
 \sigma^2_n &= \int^{\infty}_{-\infty}(x - \langle x \rangle)^2 (F \ast H ) dx \\ \nonumber
 &= \sigma^2_R + \sigma^2_r
\end{align}
where $\sigma^2_R = \int^{\infty}_{-\infty}x^2 F(x) dx$, $\sigma^2_r = \int^{\infty}_{-\infty}(x-\langle x \rangle)^2 H(x) dx$ and the functions $F$ and $H$ are normalized to unity.
From Eq.~\eqref{eq:spdineq}, it can be deduced that the width of the exact single-particle density will always be greater than the width of the COM wave function, $\sigma^2_R \leq \sigma^2_n$ and the equality is satisfied in the limit of large interaction coupling $g \to \infty$ ($\sigma^2_r \to 0$). 
Moreover, the variance of the relative motion density for an untrapped state is smaller than the untrapped case, i.e., $\sigma^2_r \lessapprox \sigma^2_{\mathrm{sol}}$, where $\sigma^2_{\mathrm{sol}}$ is an excellent approximation of the untrapped relative motion variance that becomes exact for large $N$ \cite{CalogeroDegasperis1975}. This means that the bounds for the exact single-particle density must be
\begin{equation}
  \sigma^2_R \leq \sigma^2_n \lessapprox \sigma^2_R + \sigma^2_{\mathrm{sol}}.
\end{equation}

\end{document}